\documentclass{article}

\usepackage[cp1250]{inputenc}
\usepackage{graphicx}
\usepackage{float}

\usepackage{mathtools}   
\usepackage{amsfonts}
\usepackage{url}
\usepackage{hyperref}

\begin{document}

\begin{center}
{\LARGE\centering{\bf{ \centering Cartan Connection for Schr\"{o}dinger equation. The nature of vacuum.}}} 
\end{center}


\begin{center}
\sf{Rados\l aw A. Kycia$^{1,2,a}$}
\end{center}

\medskip
\small{
\centerline{$^{1}$Masaryk Univeristy}
\centerline{Department of Mathematics and Statistics}
\centerline{Kotl\'{a}\v{r}sk\'{a} 267/2, 611 37 Brno, The Czech Republic}
\centerline{\\}
\centerline{$^{2}$Cracow University of Technology}
\centerline{Faculty of Materials Science and Physics}
\centerline{Warszawska 24, Krak\'ow, 31-155, Poland}
\centerline{\\}

\centerline{$^{a}${\tt
kycia.radoslaw@gmail.com}}
}

\begin{abstract}
\noindent

We reinterpret the Schr\"{o}dinger equation as a continuity equation in the space with the Cartan connection given by scaling Lie-B\"{a}cklund group on a specific jet space. In this space, the wave function and their gradient coordinates are treated as independent coordinates. This approach gives a full Cartan connection form a divergence-free condition. Once constructed, the connection makes it possible to investigate the geometry of the space on which this Schr\"{o}dinger-Cartan connection is constructed. This is the idea that generalizes the concepts present in de Broglie-Bohm (pilot wave) theory in a geometric way. We also present this procedure for constructing (non-uniquely) torsion-free Cartan connections for general Partial Differential Equations.
\end{abstract}

\section{Introduction}
\label{intro}

Construction of a connection for Schr\"{o}dinger equation opens up interpretation of quantum mechanics in geometrical terms. Some approaches are using the quantum connection, e.g., \cite{JosefQuantum, Josef2} or contact geometry \cite{Waldron1, Waldron2}. Moreover, the symmetry methods \cite{Vinogradow, Lychgin, Olver, KyciaJetBook} for Partial Differential Equations (PDEs) are with success applicable to Schr\"{o}dinger equations including scale transformations, e.g., \cite{Scale1, Scaling2}.

There are various notions of connection. The most general is the Ehresmann connection \cite{NaturalOperations}. The Cartan connection in physics usually appears in the context of the Einstein-Cartan theory \cite{Trautman}, and for reductive geometries, it is equivalent to the Ehresmann connection. However, the Cartan connection carries full geometric information on geometry, especially for embedded submanifolds  \cite{Sharpe}, \cite{Parabolic}, \cite{NaturalOperations}.

These results suggest the different approaches to the construction of connection for the Schr\"{o}dinger equation. We will develop a different approach that merges the scaling group and the Cartan connection in a (larger) jet space \cite{Vinogradow, Lychgin, Olver, KyciaJetBook}. As an entry point, we reinterpret Schr\"{o}dinger equation as a divergence equation in a jet space, where divergence operator is given by some Cartan connection induced by a scaling group on this jet space. Then matching the general divergence equation and the Schr\"{o}dinger equation fixes some of the Cartan connection coefficients. Unfortunately, some of them are undefined, and therefore, some additional constraint, as torsion freeness must be imposed to limit some freedom. 

The idea of some sort of scaling of the length of the object during its movement was introduced in the General Relativity context by Hermann Weyl \cite{WeylScalingRiemann}, \cite{WeylSpaceTimeMatter}. It was rejected as obviously non-physical but gave rise to the gauge principle as the scaling of phase and is base of all modern fundamental Yang-Mills type theories. In this paper, we interpret this scaling/gauge idea in the context of a change of scale of the wave function and its gradient components. This gives rise to the Cartan connection for the scaling group, which in turn, can be associated with this background, which we will call the vacuum. We will show that all these ideas can be connected in the procedure that allows us to construct a Cartan connection for Schr\"{o}dinger equation.

We sketch the basic ingredient of this procedure. Consider flat Euclidean space with 'time-space' coordinates $(t,x)$. Then we have the Schr\"{o}dinger equation (we set the Planck constant $\hbar=1$, $i=\sqrt{-1}$) for a complex-valued $L^{2}$ function $u=u(x,t)$
\begin{equation}
 i\partial_{t}u+\partial_{x}^{2}u=Vu,
 \label{Eq.Schrodinger1d}
\end{equation}
where $V=V(x,t)$ is a real-valued function called potential, which by physical assumptions is bounded from below, and it can be set\footnote{For bounded from below $V$ one can set $u\rightarrow e^{iAt}$ for $A> min_{\{t,x\}}(V(x,t))$, which results in $V\rightarrow V+A > 0$.} $V>0$. Usually $V=V(x)$, but in general $V=V(x,t)$. The additional normalization condition $\int |u(x,t)|^{2} dx=1$ is imposed for probabilistic interpretation of $|u(x,t)|^{2}dx$ as a probability measure for finding particle at $(t,x)$. This normalization, which is projection from $\mathbb{C}\setminus\{(0,0)\} \rightarrow \mathbb{C}P^{1}\cong S^2$ can be imposed in the end, so therefore we will be omitting it in what follows.

As it was pointed out in \cite{Steeb}, the equation (\ref{Eq.Schrodinger1d}) can be written as a conservation law
\begin{equation}
 d\lambda=0,
 \label{Eq.ContinuityFoLambda}
\end{equation}
for
\begin{equation}
 \lambda = in(x)u(x,t) dx - \left(n(x)\partial_{x}u(x,t)-\frac{dn(x)}{dx}\right)dt,
 \label{Eq.lambdaTensor}
\end{equation}
with the condition 
\begin{equation}
 \frac{d^{2}n(x)}{dx^2}=V(x)n(x).
 \label{Eq.VacuumEquationn(x)}
\end{equation}
The condition (\ref{Eq.VacuumEquationn(x)}) is an elliptic problem and for suitable boundary conditions the solution can be assumed to be positive\footnote{We can write a nontivial weak solution (so also a smooth one) of this equation as an unique minimizer of functional $F(n)=\int \left(\frac{1}{2} |\bigtriangledown n|^2 +\frac{1}{2} Vn^2 \right)dx > 0$ for $V>0$, i.e., $F(u)\leq F(u_{+})$, for $u_{+}:=max\{0,u\}$.}, or generally $n:\mathbb{R}^2 \rightarrow \mathbb{C}\setminus\{(0,0)\}$.

This suggest the following ansatz $u(x,t)\rightarrow n(x)u(x,t)$ in (\ref{Eq.Schrodinger1d}), that yields
\begin{equation}
 n(i\partial_{t}u+\partial_{x}^{2}u) = -2\frac{dn}{dx}\partial_{x}u +u\left(Vn - \frac{d^2n}{dx^2}\right),
 \label{Eq.1DSchrodingerNsubstitution}
\end{equation}
which can be rewritten as the system
\begin{equation}
 \left\{
 \begin{array}{c}
   i\partial_{t}u+\partial_{x}^{2}u +\frac{2}{n}\frac{dn}{dx}\partial_{x}u =0\\
   \frac{1}{n}\frac{d^{2}n(x)}{dx^2}=V(x).
 \end{array}
 \right.
 \label{Eq.VacuumSplitting1d}
\end{equation}
This decomposition, which works for linear Schr\"{o}dinger equation, has interesting interpretation. The second (elliptic) equation couples potential with scale $n$. This scale factor then couples, in the first (evolutionary) equation, to the wave function of the object under interest by the gradient term $\frac{-2}{n}\frac{dn}{dx}\partial_{x}u$. Therefore, this system of equations can be interpreted as evolution of $u$ on 'vacuum' background $n$, and thus we will be calling the sale $n$ the vacuum. In addition, due to elliptic character of equation for $n$ the background is non-casual - it is fixed once $V$ is given or, loosely speaking, 'the perturbation propagates with infinite speed'. It is worth mentioning that similar behaviour was restored in hydrodynamics by small droplet jumping on the surface of water \cite{CouderDroplets}, \cite{CouderDroplets2}, \cite{CouderDroplets3}. The droplet corresponds to isolated quantum system and the surface of water is the background on which evolution of this system happens. In the next sections we provide suitable geometric interpretation of these equations.

In order to make the link between (\ref{Eq.lambdaTensor}) and the Cartan connection consider the divergence-free condition or continuity equation \cite{Wald}, \cite{FrankelGR}
\begin{equation}
 \bigtriangledown_{a}T^{ab}=0, \quad b=1\ldots n,
 \label{Eq.ContinuityEquationGeneral}
\end{equation}
for energy-momentum tensor $T^{ab}$ in a $n$-dimensional manifold $M$ with coordinates $(x^{1},\ldots,x^{n})$ with some connection $\bigtriangledown$. In field theory it is assumed that the connection $\bigtriangledown$ arises from metric structure on $M$. However, we can reinterpret (\ref{Eq.ContinuityEquationGeneral}) as a general divergence-free equation in a space where connection is not a Riemannian connection, but arises from a Lie scaling group. We elaborate this idea further. For fixed $b$, the vector $u^{a}:=T^{ab}$ 
can be used to construct the 'flux-density' form \cite{Steeb}
\begin{equation}
 \lambda=u_{b}\star dx^{b},
\end{equation}
where $\star$ is the Hodge star induced by metric. Then (\ref{Eq.ContinuityEquationGeneral}) can be written in the form 
\begin{equation}
 d^{\bigtriangledown}\lambda=0,
\end{equation}
where $d^{\bigtriangledown}$ is covariant exterior derivative \cite{NaturalOperations}, which in flat space is usual exterior derivative in Riemannian case. This is exactly (\ref{Eq.ContinuityFoLambda}).

Now, in general case of connection for arbitrary Lie group, the equation (\ref{Eq.ContinuityEquationGeneral}) can be written in full form as
\begin{equation}
 \bigtriangledown_{i}u^{i}=\partial_{x^{i}}u^{i}+\omega^{i}_{j}(\partial_{x^{i}})u^{j} =0,
 \label{Eq.ContinuityEquationElaborated}
\end{equation}
for connection one-forms $\omega^{a}_{b}$ valued in some Lie algebra used as a gauge group. We want to stress again that in general relativity is is the Lorentz group, however we want to consider scaling group on jet space. 

We are interested in recovering connection from (\ref{Eq.ContinuityEquationElaborated}) comparing it with (\ref{Eq.1DSchrodingerNsubstitution}), however, it is difficult task to find out connection (one-forms) $\omega$ knowing only the equation (\ref{Eq.1DSchrodingerNsubstitution}). Therefore, the procedure will be non-unique in general.

The phenomenon of interaction between particle and background is explicitly present in  de Broglie-Bohm/pilot-wave approach to quantum mechanics \cite{BohmHiley}. The basic idea is to assume that $u(x,t)=\sqrt{\rho(x,t)}e^{iS(x,t)}$ for two real functions $\rho$ and $S$ interpreted as probability and phase of $u$. Upon substituting into (\ref{Eq.Schrodinger1d}) and comparing real and imaginary parts we get coupled system
\begin{equation}
 \left\{
 \begin{array}{c}
   \partial_{t}\rho + \partial_{x}(2 \rho \partial_{x} S)=0 \\
   -\partial_{t}S=(\partial_{x}S)^2 +V+Q,
 \end{array}
 \right.
\end{equation}
where $Q=-\frac{\partial_{x}^2 \sqrt{\rho}}{\sqrt{\rho}}$ is called the quantum potential. In this substitution the first equation describes conservation of the flux of probability $\rho$ that flows with the velocity $2\rho\partial_{x}S$. The second is the Hamilton-Jacobi equation with 'quantum piloting' correction $Q$ with addition to classical potential $V$. The Hamilton-Jacobi equations can be further converted to equations of motion for particle in effective $V+Q$ potential. The splitting in (\ref{Eq.VacuumSplitting1d}) is similar - $n$ is a background and $u$ represents the 'particle' moving on this background.

The paper is organized as follows: In the next section, we present the general procedure of extracting connection (or at least its part) from the arbitrary differential equation treating the equation as a divergence-free equation with connection given by a scaling group. This general procedure is then illustrated on the example of Schr\"{o}dinger equation. The paper ends with a short discussion of the results, possible implications, and philosophical questions on the nature of quantum theory.

\section{General theory}
\label{sec:1}

In this section, the general idea of recovering a Cartan connection from the partial differential equation (PDE) will be presented. This procedure is non-unique, as we will see. The next section describes the construction of specific connections for Schr\"{o}dinger equation as an illustration. Therefore it is advisable to read this section along with one of the examples of the next section.

Consider the linear differential equation, where the part of equation that contains highest order derivatives of each variable is normalized and diagonalized. We will focus on operator of second order in $x$ for simplicity, however the procedure can be applied to equations of any order. We have
\begin{equation}
 \sum_{i=1}^{n} a_{i} \partial_{i}^{2} u + A(u,\partial_{i} u)=0, 
 \label{Eq.GeneralEquation}
\end{equation}
where $A$ is a linear function in $u$ and $\partial_{i} u$ and $a_{i}=\pm 1$. We will treat this equation as a continuity equation (\ref{Eq.ContinuityEquationElaborated}) where the connection form is determined by some Lie group. 

Since continuity equation (\ref{Eq.ContinuityEquationElaborated}) does not contain each coefficient, so from our PDE we will not be able to extract the connection in unique way, however we can set all not present coefficients to zero for simplicity. However, once we set up connection for one gauge and find the gauge group $H$ then we get a geometric object that can be studied independently of PDE we started from. Here are detailed steps of the construction:

\textbf{Step 1. Introduce the jet space.} For treating (\ref{Eq.GeneralEquation}) as (\ref{Eq.ContinuityEquationElaborated}) we have to treat all derivatives up to order $1$ as variables, so we introduce the first jet space \cite{Steeb}, \cite{Vinogradow}, \cite{Lychgin}, \cite{Olver}, \cite{KyciaJetBook} on $M$, where e.g., $M=R^{n}$. For function $u$ treated as a section of $E \rightarrow M$, where, e.g., $E=M\times \mathbb{R}$, we construct $J^{1}(M)$ where local coordinates are $(x^{i}, v_{0}, u_{i})$, $i=1\ldots n$. On this space we have prolongation of $u$ to $J^{1}(M)$ given in local coordinates by $j^{1}u=(x^{i},u(x), \partial_{i}u(x))$, so $v_{0}(j^{1}u)=u$ and $v_{i}(j^{1}u)=\partial_{i}u$. Then the equation (\ref{Eq.GeneralEquation}) is
\begin{equation}
 \sum_{i=1}^{n} a_{i} \partial_{i} v_{i} + A(v_{i})=0, 
 \label{Eq.GeneralEquationJet}
\end{equation}

In general, in this step, by introducing suitable jet space, we 'flatten' $r$th order PDE to the first order $PDE$ on jet space. The number of $v_{i}$ coordinates also coincides with the number of initial/boundary data that we have to impose for the equation.

\textbf{Step 2. Select a group $G_0$.} Due to motivation in previous section we introduce a Lie-B\"{a}cklund group\cite{Steeb} on $J^{1}(M)$ that scales the differential term of (\ref{Eq.GeneralEquationJet}) by some function $a\neq 0$. We take the commutative matrix group $G$ with elements $g\in G_{0}$ of the form
\begin{equation}
 g=\left[
 \begin{array}{ccccc}
  a & 0 &  \ldots &  \ldots   & 0 \\
  b_{1} & a & \ddots  &  \ldots & 0 \\
  b_{2} & 0 & a & \ddots &   0 \\
  \vdots & \vdots & \ddots & \ddots &    0 \\
  b_{n} & 0 & \ldots & \ldots  & a
 \end{array}
\right].
\end{equation}
In general, $G_0$ does not have to be prolongation of the multiplication group $u \rightarrow a u$, that is, we does not have to fix $b_{i}=\partial_{i}a$.

Applying the group to the equation (\ref{Eq.GeneralEquationJet}) we get
\begin{equation}
 a\sum_{i=1}^{n} a_{i} \partial_{i} v_{i} +  A'(g,v,v_{i})=0, 
 \label{Eq.GeneralEquationJetGroup}
\end{equation}
where $A'$ is still linear in $v$, $v_{i}$ and depends on elements of matrix $g$ and their derivatives. 

\textbf{Step 3. Consistency condition for connection.}
The next step is to calculate all possible coefficients of connection one-form. First, introduce metric $\eta=diag(a_{1},\ldots, a_{n})$ and rewrite (\ref{Eq.GeneralEquationJetGroup}) as 
\begin{equation}
 a \eta^{ij}\partial_{i}v_{j}+A'(h,v,v_{i})=0.
\end{equation}
Then the $A'$ term is compared with (non-differential) Cartan connection part
\begin{equation}
 \frac{1}{a}  A'(h,v,v_{i}) = \omega^{i}_{j}(\partial_{x^{i}}) v_{j}.
 \label{Eq.omegaEquationGeneral}
\end{equation}
Solution for $\omega^{i}_{j}$ is non-unique since the system is not determined. However, we can set, from practical reasons, all non-occurring in (\ref{Eq.omegaEquationGeneral}) coefficients of $\omega$ to zero. Some additional conditions will appear later from demanding torsion-freeness of the curvature of this connection.

In this step we can also embed the connection $\omega$ and the group $G_0$ inside the Weyl group $G$, which is a bookkeeping device for tracing the scale $a$. The extension is
\begin{equation}
g_{0} \hookrightarrow \left[
\begin{array}{cc}
 a & 0  \\
 0 & g_{0}
\end{array}
\right],
\end{equation}
where $a$ is the diagonal element of $g_{0}\in G_{0}$ and $0$ are zero vectors of dimension $n+1$. Then the connection $\omega$ is then embedded in Weyl connection
\begin{equation}
 \omega_{W}=\left[
 \begin{array}{cc}
  \epsilon & 0 \\
  \theta  & \omega
 \end{array}
\right],
\label{Eq.WeylConnectionGeneral}
\end{equation}
where $\theta$ is a vector of dual elements to tangent base in $J^{1}(M)$, that is, $\theta = [dv_{0},\ldots, dv_{n}]^{T}$. In order to check that this is in fact the Weyl structure \cite{Sharpe} we have to show that under commutative scaling group by $b$ (that is a subgroup of $G$) the scaling part gauges as $\bar{\epsilon}=\epsilon + d\ln(b)$. This will be shown in the next step.

\textbf{Step 4. Determination of the gauge group $H$.}
We want to find such subgroup $H $ of $G$ that the differential equation (\ref{Eq.GeneralEquationJetGroup}) transforms under $h\in H$ as connection \cite{Sharpe}, \cite{Parabolic}
\begin{equation}
\omega'_{W}=Ad(h^{-1})\omega_{W}+ h^{*}\omega_{MC}.
\end{equation}
$h^{*}\omega_{MC}$ is a pullback of the Maurer-Cartan form \cite{Sharpe} for $H$ along $h$, which for matrix group is $h^{-1}dh$. Observe that the linear transformation is associated with $Ad$-transformed part and the Maured-Cartan part involves derivatives of $h$. Therefore, the natural condition is that the part of (\ref{Eq.GeneralEquationJet}) containing derivatives will transform under $h$ as continuity equation for Maurer-Cartan connection, that is,
\begin{equation}
 \eta^{ij}\partial_{i}(hv_{j}) = \lambda \eta^{ij}\partial_{i}(v_{j}) + B(h, v_j) = \lambda \eta^{ij}\partial_{i}(v_{j}) + \sum_{i} h^{-1}dh (\partial_{i})u_{i},
\end{equation}
where $\lambda$ is some arbitrary constant scale\footnote{We have a freedom to take arbitrary scale which we use here. We did not use it in the previous step for simplicity.}, and $B$ is linear in $v_{i}$. This is determining additional equation for matrix elements of $h$, so defines subgroup $H$ of $G$. Note also that the Maurer-Cartan form is determined up to scale.

To this end, for checking consistency of procedure, we have to also show that $Ad$-part also transforms in the same way as $A'(h.v_{i})$. We have for 
\begin{equation}
h=\left[\begin{array}{cc} e & 0 \\ 0 & A \end{array}\right], 
\end{equation}
that
\begin{equation}
 Ad(h^{-1})\omega_{W}=\left[\begin{array}{cc} \frac{1}{e} & 0 \\ 0 & A^{-1} \end{array}\right]\left[
 \begin{array}{cc}
  \epsilon & 0 \\
  \theta  & \omega
 \end{array}
\right]\left[\begin{array}{cc} e & 0 \\ 0 & A \end{array}\right] = \left[
 \begin{array}{cc}
  \epsilon & 0 \\
  sA^{-1}\theta  & Ad(A^{-1})\omega
 \end{array}
\right],
\label{Eq.GaugeTransformGeneral}
\end{equation}
which under change of base $\bar{\theta}=sA^{-1}\theta$ is
\begin{equation}
 Ad(h^{-1})\omega_{W} =  \left[
 \begin{array}{cc}
  \epsilon & 0 \\
  \bar{\theta}  & (Ad(A^{-1})\omega)A\frac{1}{s}
 \end{array}
\right]
\end{equation}
Since group $G$ is commutative so $Ad(A^{-1})\omega = \omega$ and we are left with lower-right block of the form $ \frac{1}{s} \omega A$, which is exactly the same operation that is made on $A'(v_{i})$ of (\ref{Eq.GeneralEquationJet}) under transformation $v_{i}\rightarrow hv_{i}$. This shows that the gauge transformations on constructed connection are consistent with the same operations on differential equation.

Note also that the $(1,1)$ block of (\ref{Eq.GaugeTransformGeneral})  is $\bar{\epsilon}=\epsilon + d\ln(e)$, therefore, the defining condition for existence of Weyl structure is indeed fulfilled.

\textbf{Step 5. Torsion-free condition.}
The curvature of (\ref{Eq.WeylConnectionGeneral}) is \cite{Sharpe}, \cite{Parabolic}, \cite{NaturalOperations}
\begin{equation}
 \Omega_{W}:=d\omega_{W}+\omega_{W} \wedge \omega_{W} = \left[
 \begin{array}{cc}
  d\epsilon  & 0 \\
  d\theta + \theta \wedge \epsilon + \omega \wedge \theta & d\omega +\omega \wedge \omega
 \end{array}
\right],
\end{equation}
and the vanishing torsion condition is 
\begin{equation}
  d\theta + \theta \wedge \epsilon + \omega \wedge \theta = 0.
\end{equation}
This condition imposes additional constraints on connection coefficients.

From the physical point of view, the vanishing torsion is a reasonable condition, which means that we require commutativity of parallel transport in jet space of solution. Since the initial vector in jet space is determined by initial data for the equation, so the torsion-free condition forces that the initial data propagates in the same way along any path on $M$.

Please also note that the torsion-freeness can be used to determine Weyl structure $\epsilon$, since it is a $1$-form.

\section{Application to Schr\"{o}dinger equation}
In this section, we construct a Cartan connection using the method presented above. First, for clarity, we present the construction in $1+1$ dimension and then generalize construction to $1+n$ case.

\subsection{$1+1$ dimensional case}
We will consider the following one-dimensional Schr\"{o}dinger equation
\begin{equation}
 i\partial_{t} u + \partial_{x}^{2}u -Vu =0,
\end{equation}
for $V=V(x,t)$ bounded from below. Underlying manifold is $M=\mathbb{R}^{2}$ and $u$ can be considered as a section of $M \times \mathbb{C}^2 \rightarrow M$. We will follow the steps described in the previous section.

\textbf{Step 1.} The order of derivatives is $(1,2)$ in $(t,x)$. Therefore, it is sufficient to introduce $J^{1}(M)$ with coordinates $(t,x,v,v_{1})$, which on holonomic sections $j^{1}u$ are $v(j^{1}u)=u$, and $v_{1}(j^{1}u)=\partial_{x}u$. On $J^{1}(M)$ there is also a coordinate associated with $\partial_{t}u$, however we do not need it and we omit it. This can be seen as a projection from $J^{1}$ to subspace where there is no 'time-derivative' coordinate. Note also that we can leave $i$ factor next to derivative, which is associated with change $t \rightarrow t/i$, or we can take $iv$ variable instead of $v$, however this introduces $i$ in $v_{1}$.

The equation in these coordinates is
\begin{equation}
 i\partial_{t}v+ \partial_{x}v_{1}-Vv=0,
 \label{Eq.1DSchrodinger}
\end{equation}
which is of first order.

\textbf{Step 2.} The group $G_{0}$ consists of $2\times 2$ matrices $g$ of the form
\begin{equation}
 g=\left[ 
 \begin{array}{cc}
  a & 0\\
  b & a
 \end{array}
\right].
\end{equation}
Under this element the equation transforms to 
\begin{equation}
 i\partial_{t}v+\partial_{x}v + v \frac{1}{a} (i\partial_{0}a + \partial_{x}b-Va) + v_{1} \frac{1}{a}(b+\partial_{x}a)=0.
 \label{Eq.1DSchrodingerGroupSubstitution}
\end{equation}

\textbf{Step 3.} The connection associated with $G_{0}$ has the form 
\begin{equation}
 \omega = \left[
 \begin{array}{cc}
  \alpha & 0 \\
  \beta & \alpha
 \end{array}
\right],
\end{equation}
where $\alpha = \alpha_{t}dt + \alpha_{x}dx$ and $\beta = \beta_{t}dt + \beta_{x}dx$.The continuity equation is
\begin{equation}
 i\partial_{t}v+\partial_{x}v + v(\alpha_{t}+\beta_{x})+ v_{1}\alpha_{x} =0.
 \label{Eq.1DSchrodingerConnection}
\end{equation}
Comparing (\ref{Eq.1DSchrodingerGroupSubstitution}) and (\ref{Eq.1DSchrodingerConnection}) we have system of equations for connection coefficients
\begin{equation}
 \left\{ 
 \begin{array}{l}
  \alpha_{t}+\beta_{x} = \frac{1}{a} (\partial_{x}b-Va) \\
  \alpha_{x} = \frac{1}{a}(b+\partial_{x}a)
 \end{array}
\right.
\label{Eq.1DSchrodingerAlphaBetaEquation}
\end{equation}
Note indeterminism of this system - we have the following symmetry
\begin{equation}
 \left\{
 \begin{array}{c}
  \alpha_{t} \rightarrow  \alpha_{t} + f(x,t)dt \\
  \beta_{x} \rightarrow \beta_{x} - f(x,t) dx\\
  \beta_{t} = g(x,t)dt,
 \end{array}
 \right.
\end{equation}
for some arbitrary functions $f$ and $g$. We can choose, e.g, $\alpha_{t}=\frac{1}{a} i\partial_{0}a dt$ and $\beta_{t}=0$. In this way we have determined non-uniquely a connection matrix $\omega$.

We can also embed it in Weyl structure. The Weyl connection is
\begin{equation}
 \omega_{W}=\left[
 \begin{array}{ccc}
  \epsilon & 0 & 0 \\
  dv  & \frac{1}{a}(b+\partial_{x}a)dx +\frac{1}{a} i\partial_{0}a dt  & 0 \\
  dv_{1} & \frac{1}{a} (\partial_{x}b-Va) dx & \frac{1}{a}(b+\partial_{x}a)dx +\frac{1}{a} i\partial_{0}a dt
 \end{array}
\right],
\label{Eq.1DSchrodingerWeylConnection}
\end{equation}
for some $\epsilon$ to be determined later.

\textbf{Step 4.} We will now construct a subgroup $H_{0}$ of $G_{0}$ with element $h$ of the form
\begin{equation}
 h=\left[ 
 \begin{array}{cc}
  e & 0 \\
  f & e
 \end{array}
\right].
\end{equation}
The Maurer-Cartan form for this element is
\begin{equation}
 h^{*}\omega_{MC}=h^{-1}dh = \left[
 \begin{array}{cc}
  d\ln(e) & 0\\
  d\frac{f}{e} & d\ln(e)
 \end{array}
\right]
\label{Eq.1DSchrodingerMaurerCartan}
\end{equation}

We want to associate in part containing the highest order derivatives in the equation with (\ref{Eq.1DSchrodingerMaurerCartan}). From (\ref{Eq.1DSchrodinger}) we get after using $h$
\begin{equation}
 i\partial_{t}(hv)+\partial_{x}(hv_{1})=i\partial_{t}v+\partial_{x}v_{1} + v \frac{1}{e} (i\partial_{t}e + \partial_{x}f) + v_{1} \frac{1}{e}(f+\partial_{x}e)=0,
\end{equation}
and from the continuity equation for connection given by (\ref{Eq.1DSchrodingerMaurerCartan}) we obtain
\begin{equation}
 i\partial_{t}v+\partial_{x}v_{1} + v (\frac{1}{e} \partial_{t} e + \partial_{x}(f/e) ) + v_{1} \frac{1}{e}\partial_{x}e =0.
\end{equation}
The coefficients at $v$ and $v_{1}$ should be proportional\footnote{The linear equation has symmetry of scaling by nonzero constant factor $\lambda$, which is a special case of group we consider.} and therefore we get the system
\begin{equation}
 \left\{ 
 \begin{array}{c}
  \frac{1}{e} (i\partial_{t}e + \partial_{x}f) = \lambda (\frac{1}{e} \partial_{t} e + \partial_{x}(f/e) ) \\
   \frac{1}{e}(f+\partial_{x}e) = \lambda \frac{1}{e}\partial_{x}e
 \end{array}
 \right.
\end{equation}
These system can be rewritten as
\begin{equation}
 \left\{ 
 \begin{array}{c}
  f = (\lambda-1)\partial_{x}e \\
  (i-\lambda)\partial_{t} \ln(e) -(\lambda -1 )(\partial_{x}\ln(e))^{2} = \partial_{x}^{2}\ln(e)
 \end{array}
\right.
\end{equation}
which is a constraint for group manifold $H$. Note that for $\lambda=1$ we get $f=0$ and $H_{0}$ consist of diagonal matrices.

\textbf{Step 5.} Starting from (\ref{Eq.1DSchrodingerWeylConnection}) and using the fact that we use holonomic coordinate system $\theta=[dv, dv_{1}]^{T}$, that is $d\theta=0$, we get vanishing torsion condition in the form
\begin{equation}
 0=\theta\wedge \epsilon + \omega \wedge \theta = 
 \left[
 \begin{array}{c}
 \epsilon \\
 \epsilon
 \end{array}
\right] + 
\left[
\begin{array}{cc}
 \alpha & 0 \\
 \beta & \alpha
\end{array}
\right] 
\left[ 
\begin{array}{c}
 1 \\
 1
\end{array}
 \right].
\end{equation}
This gives $\epsilon =-\alpha=-(\partial_{x}\ln(a) dx +\frac{1}{a} i\partial_{0}a dt)$ and $\beta = 0$. Therefore, torsion-free condition is
\begin{equation}
 0= \frac{1}{a} (\partial_{x}b-Va).
\end{equation}
It is understand that once we choose $a$, then this equation determines $b$, or vice versa.

We can consider special cases, when the group $G_{0}$ is a prolongation to $J^{1}(M)$ of the multiplication group $u\rightarrow au$, i.e., $b=\partial_{x}a$. We get then
\begin{equation}
 \frac{1}{a}\partial_{x}^{2}a-V =0,
\end{equation}
which is exactly the condition from (\ref{Eq.VacuumSplitting1d}) for $a=n$. This suggest that the form of (\ref{Eq.VacuumSplitting1d}) has fundamental geometric meaning. However this condition changes, when we choose differently $\alpha_{t}$ and $\beta_{x}$ tu fulfil the equation (\ref{Eq.1DSchrodingerAlphaBetaEquation}).

\subsection{$1+n$ dimensional case}
The $1+n$ ($n>0$) case is a straightforward generalization of previous subsection.

\textbf{Step 1.} We have $M=\mathbb{R}^{1+n}$ with coordinates $(t,x^{1},\ldots, x^{n})$. Then for the functions $u:M\rightarrow \mathbb{C}$, which can be seen as sections of $M\times \mathbb{C} \rightarrow M$, we construct jet bundle $J^{1}(M)$ with coordinates$(t,x, v,v_{1},\ldots, v_{n})$, where we have $v_{i}(j^{1}u)=\partial_{x^{i}}u$. As before, the coordinate associated with $\partial_{t}u$ is excluded by the projection. 

The Schr\"{o}dinger equation is
\begin{equation}
 i\partial_{t}v+ \sum_{i=1}^{n}\partial_{i}v_{i} -Vv=0.
 \label{Eq.SchrodingerN}
\end{equation}

\textbf{Step 2.} The group $G_{0}$ contains matrices of the form
\begin{equation}
 g=\left[
 \begin{array}{ccccc}
  a & 0 &  \ldots &  \ldots   & 0 \\
  b_{1} & a & \ddots  &  \ldots & 0 \\
  b_{2} & 0 & a & \ddots &   0 \\
  \vdots & \vdots & \ddots & \ddots &    0 \\
  b_{n} & 0 & \ldots & \ldots  & a
 \end{array}
\right],
\end{equation}
with connection one-form
\begin{equation}
 \omega=\left[
 \begin{array}{ccccc}
  \alpha & 0 &  \ldots &  \ldots   & 0 \\
  \beta_{1} & \alpha & \ddots  &  \ldots & 0 \\
  \beta_{2} & 0 & \alpha & \ddots &   0 \\
  \vdots & \vdots & \ddots & \ddots &    0 \\
  \beta_{n} & 0 & \ldots & \ldots  & \alpha
 \end{array}
\right],
\label{Eq.ConnectionGeneral}
\end{equation}
with $\alpha = \alpha_{t}dt + \sum_{i=1}^{n}\alpha_{i}dx^{i}$ and $b_{i}=\beta_{t}dt + \sum_{j=1}^{n}\beta_{ij}dx^{j}$.

\textbf{Step 3.} The equation after transformation by $g$ is 
\begin{equation}
  i\partial_{t}v+ \sum_{i=1}^{n}\partial_{i}v_{i} + v\frac{1}{a}(i\partial_{t}a +\sum_{i=1}^{n} \partial_{i}b_{i} - Va) + \sum_{i=1}^{n}v_{i}\frac{1}{a}(b_{i}+\partial_{i}a) =0,
 \label{Eq.SchrodingerNGroup}
\end{equation}
and the continuity equation for connection (\ref{Eq.ConnectionGeneral}) is 
\begin{equation}
 i\partial_{t}v+ \sum_{i=1}^{n}\partial_{i}v_{i} + v\left(\alpha_{t} + \sum_{i=1}^{n}\beta_{ii}\right) + \sum_{i=1}^{n} v_{i} \alpha_{i} =0.
\end{equation}
Comparing coefficients at $v$ and $v_{i}$ we obtain an undetermined system of equations for connection
\begin{equation}
 \left\{ 
 \begin{array}{c}
  \left(\alpha_{t} + \sum_{i=1}^{n}\beta_{ii}\right) = \frac{1}{a}(i\partial_{t}a +\sum_{i=1}^{n} \partial_{i}b_{i} - Va) \\
  \alpha_{i} = \frac{1}{a}(b_{i}+\partial_{i}a).
 \end{array}
 \right.
\end{equation}
Here the choice of connection coefficient is also non-unique. Therefore, we will do it in analogical way as for $1+1$ case, and symmetrize potential $V$ among $\beta$ coefficients, i.e.,
\begin{equation}
\left\{
 \begin{array}{l}
  \alpha = \frac{1}{a}i\partial_{t}a dt + \sum_{i=1}^{n} \frac{1}{a}(b_{i}+\partial_{i}a) dx^{i}\\ \\
  \beta_{i} =  \frac{1}{a}(\partial_{i}b_{i} - \frac{1}{n} Va) dx^{i} \quad (\mathrm{no} \quad \mathrm{summation}).
 \end{array}
 \right.
\end{equation}

The embedding of connection into the Weyl connection is as before
\begin{equation}
 \omega_{W} = \left[
 \begin{array}{cc}
  \epsilon & 0 \\
  \theta & \omega
 \end{array}
\right],
\end{equation}
where $\theta = [dv, dv_{1},\ldots, dv_{n}]^{T}$, and the Weyl structure $\epsilon$ will be determined later by torsion-free condition.

\textbf{Step 4.} We will now look for $H_{0}$ subgroup of $G_{0}$ that we match with by differential part of the equation with the Maurer-Cartan form for $H_{0}$. Element $h\in H_{0}$ is of the form
\begin{equation}
 h=\left[
 \begin{array}{ccccc}
  e & 0 &  \ldots &  \ldots   & 0 \\
  f_{1} & e & \ddots  &  \ldots & 0 \\
  f_{2} & 0 & e & \ddots &   0 \\
  \vdots & \vdots & \ddots & \ddots &    0 \\
  f_{n} & 0 & \ldots & \ldots  & e
 \end{array}
\right],
\end{equation}
and the Maurer-Cartan form for $h$ is
\begin{equation}
 h^{*}\omega_{MC}= h^{-1}dh =\left[
 \begin{array}{ccccc}
  d\ln(e) & 0 &  \ldots &  \ldots   & 0 \\
  d \frac{f_{1}}{e}  & d\ln(e) & \ddots  &  \ldots & 0 \\
  d \frac{f_{2}}{e} & 0 & d\ln(e) & \ddots &   0 \\
  \vdots & \vdots & \ddots & \ddots &    0 \\
  d \frac{f_{n}}{e} & 0 & \ldots & \ldots  & d\ln(e)
 \end{array}
\right].
\end{equation}

The differential part of (\ref{Eq.SchrodingerN}) under $h$ is changed to 
\begin{equation}
  i\partial_{t}v+ \sum_{i=1}^{n}\partial_{i}v_{i} + v\frac{1}{e}(i\partial_{t}e +\sum_{i=1}^{n} \partial_{i}f_{i}) + \sum_{i=1}^{n}v_{i}\frac{1}{e}(f_{i}+\partial_{i}e)=0,
 \label{Eq.SchrodingerNGroup}
\end{equation}
which we compare with continuity equation for the Maurer-Cartan form of $H_{0}$, that is,
\begin{equation}
 i\partial_{t}v+ \sum_{i=1}^{n}\partial_{i}v_{i} + v\left(\partial_{t}\ln(e) + \sum_{i=1}^{n}\partial_{i}(f_{i}/e) \right) + \sum_{i=1}^{n} v_{i} \partial_{i} \ln(e) =0.
\end{equation}
The coefficients at $v$, $v_{i}$ of both equations should be proportional by some constant $\lambda$, and therefore we have matching conditions
\begin{equation}
 \left\{ 
 \begin{array}{c}
  f_{i} = (\lambda-1) \partial_{i}e \\
  (i-\lambda)\partial_{t}\ln(e) -\sum_{i=1}^{n}(\partial_{i}\ln(e))^{2}=(\lambda-1)\sum_{i=1}^{n} \partial_{i}^{2}\ln(e),
 \end{array}
\right.
\end{equation}
which gives conditions defining $H_{0}$.

\textbf{Step 5.}
The final step is to impose torsion-free condition. Since we are using holonomic coordinates, so $d\theta=0$, and the condition is
\begin{equation}
 0=\theta\wedge \epsilon + \omega \wedge \theta = 
 \left[
 \begin{array}{c}
 \epsilon \\
 \vdots \\
 \vdots \\
 \vdots \\
 \epsilon
 \end{array}
\right] +  \left[
 \begin{array}{ccccc}
  \alpha & 0 &  \ldots &  \ldots   & 0 \\
  \beta_{1} & \alpha & \ddots  &  \ldots & 0 \\
  \beta_{2} & 0 & \alpha & \ddots &   0 \\
  \vdots & \vdots & \ddots & \ddots &    0 \\
  \beta_{n} & 0 & \ldots & \ldots  & \alpha
 \end{array}
\right]
\left[ 
\begin{array}{c}
 1 \\
 \vdots \\
 \vdots \\
 \vdots \\
 1
\end{array}
 \right].
\end{equation}

That gives $\epsilon = -\alpha$ and $\beta_{i}=0$. This means that for our specific choice of connection we have
\begin{equation}
 \partial_{i}b_{i} - \frac{1}{n} Va =0, \quad i=1,\ldots,n, 
\end{equation}
or
\begin{equation}
 \sum_{i=1}^{n} \partial_{i}b_{i} = Va.
\end{equation}
This can be treated as equation for $b$ when $a$ is some selected function or vice verso.

Again, for $G_{0}$ that is lift of multiplication group on $u$, that is $b_{i}=\partial_{i}a$, we get condition in the form 
\begin{equation}
 \bigtriangleup a = Va.
\end{equation}
This is $n$-dimensional version of (\ref{Eq.VacuumEquationn(x)}). Again, different choices for $\alpha$ and $\beta$ results in change of this condition.

\subsection{Other equations}
Concluding this section, we also comment on other PDEs.

The results of this section easily extend to the diffusion equation by changing $i\rightarrow 1$. The Schr\"{o}dinger equation is vital due to motivation and physical context, however, the same procedure can be applied to other PDEs.

\section{Conclusions}
The above construction of the Cartan connection for the scaling group of wave functions shows that the Schr\"{o}dinger equation is of geometric type. Since it is associated with the continuity equation, therefore it defines only some of the connection components. The other components must be fixed otherwise. This shows that the Schr\"{o}dinger equation has big freedom of selecting the connection one-form, and therefore, the freedom to be geometrized. The torsion-free condition for the curvature of the connection imposes additional conditions on connection and makes the connection to belonging to the diagonal subgroup of scalings on jet space.

The above geometric approach is present in the first jet space where components of the vector are the wave function and its gradient. However, knowing the Cartan connection, we can also ask how arbitrary vector in jet space, that apart of the wave function has other components that are not necessary gradient components of the wave function, evolves. Therefore, by the construction described in the paper, we significantly extended degrees of freedom of quantum mechanics.

As it was mentioned in the Introduction, the scaling function (sometimes non-normalizable) appears in many places in quantum mechanics. The approach presented in this paper assigns to this scaling a geometric object that can be associated with the background on which quantum evolution takes place. We believe that it is the first step to a better understanding of the nature of quantum vacuum and construct powerful non-perturbative geometric methods for solving quantum-mechanical problems.

The procedure of constructing the Cartan connection from Schr\"{o}dinger equation is non-unique because the equation is treated as a continuity equation, which is some kind of 'divergence-free' condition of some object. Therefore, this approach suggests that some additional information must be imposed if quantum mechanics from this viewpoint should be considered as a geometric theory of Cartan connection. This raises a serious question on completeness of quantum theory described by Schr\"{o}dinger equation. In this approach, the same problem appears if we consider general PDE and try to treat it as a continuity equation. From a philosophical point of view, this lack of additional input perhaps may be attributed to the fact that the continuity equation describes how we can manipulate of a given quantity of 'material' to preserve it, however, it does not describe what kind of 'material' on which we manipulate. Therefore to describe of nature of this 'material' additional set of laws has to be provided.

From a mathematical point of view, the construction of the Cartan connection for PDE was provided. The procedure is non-unique, however, once performed, it can give a geometric interpretation of the equation.


\section*{Acknowledgments}
These ideas in the paper grew for many years during discussions. I would like to thanks Andrew Waldron for inspiring discussion on gauge theories and quantum mechanics during 38th Winter School of Geometry and Physics in Srni, to Josef Silhan and Jan Slov\'{a}k for support and for showing me a beautiful world of differential geometry of Cartan connection, and to Jarek Duda for discussion on nature of space and time. Last but not least, I would like to thanks Anatoliy K. Prykarpatsky for continuous motivation, and Rikard von Unge and Klaus Bering Larsen for discussion on the nature of the quantum world. 

The author was supported by the GACR grant GA19-06357S, the grant 8J20DE004 of Ministry of Education, Youth and Sports of the CR, and Masaryk University grant MUNI/A/0885/2019. I would like to acknowledge COST CA18223 action for support.

\appendix


\end{document}